# A visible - infrared compatible camouflage photonic crystal with enhanced emission in 5~8 *μ*m

Saichao Dang, Hong Ye*

Department of Thermal Science and Energy Engineering, University of Science and Technology of China, Hefei 230027, People's Republic of China.

*Correspondence to: H. Ye (E-mail: hye@ustc.edu.cn)

**Abstract**: Because of surface structural constraint and thermal management requirement, visible - infrared compatible camouflage is still a great challenge. In this study, we introduce a 2D periodic aperture array into ZnO/Ag/ZnO film to realize visible-infrared compatible camouflage with a performance of thermal management by utilizing the extraordinary optical transmission in a dielectric/metal/dielectric (D/M/D) structure. Because of the high visible transmittance of the D/M/D structure, when applied on a visible camouflage coating, the beneath coating can be observed, realizing arbitrary visible camouflage. Due to the perforated Ag layer, both low emittances in 3~5 *μ*m, 8~14 *μ*m for infrared camouflage and high emittance in 5~8 *μ*m for heat dissipation by radiation are achieved theoretically and experimentally. The fabricated photonic crystal exhibits high-temperature infrared camouflage in two atmospheric windows. With the same heating power of 0.40 W/cm$^2$, this photonic crystal is 12.2 °C cooler than a sample with a low-emittance surface. The proposed visible - infrared compatible camouflage photonic crystal with the performance of thermal management provides a guideline on coordinated control of light and heat, indicating a potential application in energy & thermal technologies.

# Introduction

With the development of detection technologies, visible-infrared compatible camouflage (VICC) has become a great challenge because of the mismatched surface structural requirements of two different camouflages (i.e., visible: color control, infrared: radiation signature modulation) and thermal management issue [1-4]. Adopting a coating exhibiting a similar color to a background is a common method to realize visible (380~780 nm) camouflage [5, 6]. Unfortunately, this coating can hardly have a low infrared emittance in two atmospheric windows (i.e., 3~5 $\mu$m and 8~14 $\mu$m), which is required by the infrared camouflage [7-10]. Similarly, an infrared coating with a specific color shows a limited performance of visible camouflage [10]. What's more, the low emittance over the whole infrared range of an infrared camouflage suppresses the radiative heat transfer [1-4], causing severe heat instability, especially for a high-temperature target (e.g., converging nozzles of aircrafts (~950 K) [11] and funnels of naval ships (~680 K) [12]). As the thermal radiation of a high temperature target in 5~8 $\mu$m accounts for a large amount (e.g., 32.33% for a 680 K blackbody), the target's temperature can be decreased because of a high emittance in this band, further weakening its infrared signal intensity [1-3]. Although several wavelength-selective infrared camouflage structures with a high emission in 5~8 $\mu$m are proposed based on multilayer [1, 2] or photonic structures [3, 4], their visible camouflage performances are limited because of the drab color and the infrared camouflage performances are seldom demonstrated in both 3~5 $\mu$m and 8~14 $\mu$m. In this work, a novel photonic crystal is proposed for general VICC based on a dielectric/metal/dielectric (D/M/D) film perforated with 2D periodic array of apertures.

A D/M/D film can realize a high visible transmittance and a high infrared reflectance (low emittance) and has been applied as a transparent electrode or heat mirror due to the infrared reflection of the middle metal layer and the suppression of visible reflection by the two aside dielectric layers [13-18]. When applied on a visible camouflage coating, a D/M/D film can make the beneath arbitrary colored coating observed due to its high visible transmittance, and realize infrared camouflage [9] simultaneously because of its low emittance over the total mid-infrared range [9, 13-18]. To prepare a D/M/D film of a high emission in 5~8 $\mu$m, we will introduce a 2D periodic aperture array into this film to utilize the extraordinary optical transmission (EOT). The EOT has been a very active research field [21-27] since its discovery through nanoscale apertures in metallic films in 1998 [19,20], leading to its application in sensing [28-33], spectroscopic devices [34, 35], perfect absorbers [36, 37], color filters [38-41], metamaterials [42-46], lenses [47], optical trapping [48-50], enhancement of nonlinear effects [51-53], etc. Thanks to the periodic subwavelength aperture array, coupled effect occurs between the incident electromagnetic wave and aperture array, leading to a wavelength-selective transmission at frequencies below the first onset of diffraction with higher efficiencies than that expected for subwavelength apertures [19-28]. The resonant wavelengths can be tuned by varying the periodicity and the symmetry of an aperture array or the periodic features surrounding individual apertures [20-25]. By introducing infrared subwavelength array of apertures into a D/M/D film, it is expected that a high emittance in 5~8 $\mu$m can be realized to enhance heat dissipation by radiation, with the VICC performance retained simultaneously.

In this work, a novel VICC photonic crystal, i.e., a ZnO/Ag/ZnO film perforated with a 2D periodic array of apertures, is proposed and demonstrated with enhanced heat dissipation by radiation. Its radiative properties in visible and infrared ranges are measured. Those properties are also simulated by the finite difference time domain (FDTD) method. What's more, both the theoretical and experimental performances of the VICC and thermal management of the fabricated photonic crystal are demonstrated.

## Results and Discussion

### Scheme of a VICC photonic crystal for a high temperature target

The scheme of a VICC photonic crystal for a high temperature target, which combines a thermal insulator, a visible camouflage coating and a photonic crystal, is shown in Figure 1a. The thermal insulator covers the high temperature target to reduce the conduction heat transfer from the high temperature target. A visible camouflage coating is covered on the insulator. The transparent photonic crystal is placed at the surface of the whole structure to make the beneath colored coating observed. Ideally, zero emittances (Figure 1b) in two atmospheric windows (3~5 $\mu$m and 8~14 $\mu$m) reduce the infrared signal intensity and the unity emittance in 5~8 $\mu$m (see the spectrum in Figure 1b or Table 1) enhances the heat dissipation by radiation. For a high temperature (e.g., 680 K) object with a high infrared emittance, a large amount of energy (32.33% for a 680 K blackbody) can be released through 5~8 $\mu$m (Figure 1b), which can cool the object and weaken its infrared signal further. To realize high-temperature VICC, a ZnO/Ag/ZnO film perforated with 2D periodic array of apertures (Figure 1c) is demonstrated. The triple-layered ZnO/Ag/ZnO shows a high visible transmittance and a low emittance in the total mid-infrared range. Because a D/M/D film with a 2D periodic aperture array can show an extraordinary transmission, the thermal emission from the beneath coating can be selective transmitted. Thus, it is expected that, with proper design, this novel photonic crystal can realize VICC and heat dissipation by radiation for a high temperature target, simultaneously.

To demonstrate the performances of infrared camouflage and thermal management of a VICC photonic crystal, an energy balance model as shown in Figure 1a is applied to calculate the surface temperature of the whole structure ($T_s$). The surface of the high-temperature target (area $A$) is a boundary with a temperature of $T_h$. The thermal insulator is modeled with a thermal conductivity of 0.04 W/(m·K) and a thickness of 1 cm (silica aerogel [1]), assuming the conductivity is temperature independent. The thermal resistances of the visible camouflage coating and VICC photonic crystal are neglected. The heat loss due to convection of the top surface can be calculated by

$$P_{conv} = h_{conv} A (T_s - T_{atm}) \qquad (1)$$

with the ambient temperature ($T_{atm}$) being 300 K and $h_{conv}$ being the convection heat transfer coefficient (assuming natural convection with $h_{conv}$=10 W/(m$^2$·K) in this work). According to Planck's law, the integral radiation powers of the top surface in three bands (shown in Figure 1a) can be calculated. The surroundings is assumed to be a 300 K blackbody. According to the Kirchoff's law (absorbance=emittance), the absorbed radiation by the structure from the surroundings can be obtained through the emittance in Table 1. Assuming 1D heat transfer, the surface temperature ($T_s$) can be calculated.

As shown in Figure 2a, the surface temperatures of the two structures (I and II) with a broadband zero-emittance and ideal VICC for high temperature target (Table 1), respectively, are calculated without considering the thermal stability of each component. With a thermal insulator, both structures can reduce the temperature and the structure II shows a larger temperature reduction. As the target temperature increases, the thermal radiation in 5~8 μm becomes more significant, leading to an increased surface temperature reduction. At the highest target temperature of 2000 K, the surface temperature reductions of insulator with structure I and insulator with structure II are 1334 K, and 1423 K, respectively. Combining thermal insulator and radiation in 5~8 μm can significantly increase the applicable object temperature range for infrared camouflage.

To investigate the contribution of radiation in 5~8 μm to the infrared camouflage, two low-emittance structures (i.e., III and IV) are assumed with their spectral emittance given in Table 1. The infrared signal intensity of an object is estimated by integrating the infrared radiance in an atmospheric window (i.e., $\lambda_1 \sim \lambda_2$):

$$I = 10\log_{10}\left(\int_{\lambda_1}^{\lambda_2} \varepsilon(\lambda) I_{BB}(\lambda,T) d\lambda\right) \quad (2)$$

where $I_{BB}(\lambda,T)$ is the spectral radiance of a blackbody at temperature $T$ and $\varepsilon(\lambda)$ is the surface emittance. For comparison, the infrared signal intensity of the high temperature blackbody is regarded as the baseline (0 dB). The signal reduction within two atmospheric windows are shown in Figure 2b and 2c. In the case of only thermal insulator with a blackbody radiation, the surface temperature is effectively reduced, further causing a reduction of infrared signal intensity. Within 3~5 μm window as shown in Figure 2a, the signal intensity is reduced by more than 10 dB (blue area) when the temperature is higher than 400 K. Assuming the surface emittance is reduced to $\varepsilon$ =0.05, the signal is further reduced by approximately 7 dB (purple area). While within 8~14 μm window as shown in Figure 2c, the signal intensity is reduced by more than 4 dB when the temperature is higher than 400 K. The signal intensity can be reduced by 10 dB with the low emittance. With heat dissipation by radiation in 5~8 μm, the surface temperature further decreases, and its contribution to infrared camouflage is more significant when the object temperature is higher (orange area in two figures). When the target temperature is 1000 K, the infrared signal intensity can be reduced by 33 dB and 22 dB within 3~5 μm and 8~14μm windows, respectively, with simultaneous thermal insulation, low emittance in the atmospheric windows and heat dissipation by radiation in 5~8 μm, demonstrating the applicability of high-temperature camouflage via effective thermal management.

**Samples and the radiative properties**

Figure 3 shows the prepared sample, i.e., photonic crystal on a silica substrate (PCS) as well as its radiative properties. The directional-hemispherical transmittance ($T_{dh}$) and the directional-specular transmittance ($T_{ds}$) of the sample in visible band are shown in Figure 3a. High visible transmission is realized, indicating the beneath stuff can be observed clearly as shown in Figure 3b. This high transparency is mainly caused by Fabry–Pérot resonance effect [13-18] in the triple-layed ZnO/Ag/ZnO structure (see SEM in Figure 3a). The reflected waves from the interfaces will superpose with one another to form a destructive reflective interference, leading to a high transmission. The light is considered normally incident, under which the specular radiative properties equal hemispherical ones for a multilayer structure. However, the light will be diffracted by a periodic surface (Figure 3c) with multiple diffraction orders in Fourier space [54-58]. As shown in Figure 3a, the measured $T_{dh}$ or $T_{ds}$ is lower than that of the corresponding simulated one because light will be scattered by the rough edge of apertures (as shown in Figure 3d), causing a reduction of the transmittance. According to measurements in Figure 3a, the $T_{dh}$ can reach 88% and $T_{ds}$ can reach 82%. Due to the high transmittance, visible camouflage can be realized by applying this structure on a visible camouflage coating.

To investigate the mechanism of EOT occurred in the infrared band, a free-standing photonic crystal (FSPC) is also simulated in this study. The radiative properties of an FSPC and PCS in infrared (3~ 14 $\mu$m) are shown in Figure 3e and 3f, respectively. According to Bloch-Floquet condition [54-59], the minimum supported vector by a periodic structure with period $P$ is $2\pi/P$. For this considered square array with period $P$=5.5 $\mu$m under normal incident situation, there will be only zero order diffracted wave when the wavelength is larger than $P$ that can propagate in air (reflection region), which can be observed from the simulations. At a shorter wavelength than $P$, the incident wave will be diffracted in high orders, indicating directional-hemispherical reflectance ($R_{dh}$) or transmittance ($T_{dh}$) higher than the specular one ($R_{ds}$ or $T_{ds}$) as shown in Figure 3e and 3f. As shown in Figure 3e, due to the reflection of Ag layer [8, 9], high reflectance (i.e., low emittance) can be observed in 3~5 $\mu$m and 8~14 $\mu$m, indicating a great camouflage performance in these bands. The transmission maximum within 5~8 $\mu$m indicates that the emission from beneath stuff can get transmitted for heat dissipation by radiation. As shown in Figure 3e, a transmission minimum shows up at a wavelength close to the period $P$, coinciding with the Rayleigh's condition [22, 58, 60], and an accompanying transmission maximum shows up at a longer wavelength, revealing its connection to Wood's anomalies [22, 61]. Reflection peaks occur at the Rayleigh's condition that beam becomes grazing while the accompanying transmission peaks are caused by the surface plasmon polaritons [22, 58,60]. The Rayleigh-Wood's anomalies can be observed from the normalized electric fields at the cross-section of a single unit, as shown in Figure 4. At λ=$P$ (Figure 4a), the $E_x$ distribution means high reflection and the $E_z$ distribution shows a delocalized surface wave, matching the Rayleigh's condition. At the accompanying transmission maximum wavelength λ=6.06 $\mu$m (Figure 4b), the strong confined electric field can be observed, indicating a resonant interaction of the incident wave with the surface plasmons on both surfaces of the Ag layer, causing a maximum transmission. At λ=14 $\mu$m, the observed high reflection is caused by the intrinsic high permittivity of Ag (Figure 4c).

In the present work, the photonic crystal was prepared on a silica substrate for demonstration. Silica is transparent in the visible band while it has strong phonon–polariton resonance within the infrared band. Both the simulated and measured transmittances of PCS sample are 0 in infrared band, which is different from that of FSPC. As shown in Figure 3e and 3f, the reflectance of PCS sample is lower than that of FSPC in two atmospheric windows, especially in 8~14 μm. Thus, the FSPC shows a better performance of infrared camouflage. As shown in Figure 3f, the simulated $R_{dh}$ has a similar trend with the measured one especially at a wavelength larger than 6 μm. The valley of the simulation near 7 μm shifts to a shorter wavelength compared with the measured one. The radiative properties calculated by FDTD method are obtained by one Gaussian pulse propagating in the media with its optical parameters fitted by Lorentz model [58]. The shift for the valley near 7 μm may be caused by the fitting error for silica. From the simulated results of PCS, one can find several typical wavelengths (e.g., the reflection peaks at $\lambda = 5.5\ \mu m$ and $\lambda = 6.4\ \mu m$, and the reflection valley at $\lambda = 6.8\ \mu m$). According to the plotted electric fields at $\lambda = 5.5\ \mu m$ and $\lambda = 6.4\ \mu m$, as shown in Figure 4d and 4e, one can find the unbonded surface wave in air and silica, respectively, which match the corresponding Rayleigh's condition. From Figure 4f, the confined electric field can be observed, indicating a strong absorption (i.e., high emission). What's more, the simulated reflection at $\lambda = 14\ \mu m$ is high as shown in Figure 4g. The lowest emittances of PCS sample in 3~5 μm and 8~14 μm are 23% at $\lambda = 3\ \mu m$ and 13% at $\lambda = 8.9\ \mu m$, respectively. What's more, a high emittance of 83% at $\lambda = 7.4\ \mu m$ is obtained. In general, the PCS sample shows a high emittance in 5~8 μm for radiative cooling and low emittances in 3~5 μm and 8~14 μm for infrared camouflage.

**Experimental demonstration of VICC at high temperature**

Using a sand background, the visible camouflage of the PCS is demonstrated, as shown in Figure 5a. Due to the high visible transmittance of PCS, the beneath sand is observed clearly. This PCS can be applied on arbitrary visible camouflage coating for various visible camouflage. What's more, the PCS is put on a heating plate to demonstrate its infrared camouflage performance as shown in Figure 5b. Through two infrared cameras which work at two atmospheric windows, respectively, the infrared images and the radiation temperatures at different temperatures can be obtained, as shown in Figure 5c. For excellent infrared camouflage, an object is "invisible" under a infrared camera when its radiation temperature matches that of the background. As can be seen in the infrared images shown in Figure 5c, the prepared PCS sample is hidden into the background, while the rest part not covered by the sample is thermally detected. It should be noticed that the edge of the sample is slightly damaged, which weakens its infrared camouflage performance. The "radiation temperature" ($T_r$) recorded by an infrared camera is related to the detected power ($P$) from the surface of the sample and can be calculated through the inverse function of $P(\varepsilon_i, T)$:

$$T_r = P^{-1}(\varepsilon_i, T) \qquad (3)$$

where $\varepsilon_i$ is the default emissivity (usually $\varepsilon_i=1$) in the infrared camera; $P(\varepsilon_i, T)$ includes the radiation from the object ($P_{rad}$) and the reflection of the ambient radiation by the object ($P_{ref}$) (Figure 5b):

$$\begin{aligned}P(\varepsilon_i, T) &= P_{rad}(\varepsilon, T) + P_{ref}(\varepsilon, \varepsilon_a, T_{amb}) \\ &= C\int_{\lambda_1}^{\lambda_2}\varepsilon(\lambda)I_{BB}(\lambda, T)d\lambda + C\int_{\lambda_1}^{\lambda_2}[1-\varepsilon(\lambda)]\varepsilon_a I_{BB}(\lambda, T_{amb})d\lambda\end{aligned} \qquad (4)$$

where $\varepsilon_a$ and $T_{amb}$ is the spectral emittance and temperature of ambient; $T_s$ is the surface temperature of the sample; [$\lambda_1$, $\lambda_2$] is the working spectrum of the infrared camera; $C$ is the angle integral constant (assuming $C=1$ [2]). In the present study, the ambient temperature is 20 °C with its emittance $\varepsilon_a \approx 1$ as the surroundings can be regarded as a blackbody.

The radiation temperatures of the PCS sample at different real temperatures are measured with two infrared cameras (3~5 $\mu$m: FLIR ThermaCAM S65 and 8~14 $\mu$m: Fluke Tix600), as shown in Figure 6a. In both cases, the measured radiation temperatures ($T_r$) of the sample are much lower than its real temperatures ($T_s$), and the radiation temperatures grow up with the increasing real temperatures. When $T_s$=164.5 °C, the radiation temperatures are 99.8 °C and 68.8 °C within 3~5 $\mu$m and 8~14 $\mu$m, respectively. Extraordinary temperature reductions are realized. According to Figure 6a, one can find that the calculated result is in good agreement with the measured data. The infrared camouflage performance of the FSPC can be evaluated through calculation as well. As shown in Figure 6a, the FSPC (dash lines) shows a better infrared camouflage performance than that of PCS.

**Experimental demonstration of thermal management performance**

The thermal management performance of the PCS is demonstrated by comparing the temperatures of an equal-sized silica with an on-top Al foil film. The ceramic heating plate is applied with its energy provided by a DC power supply, and the temperature is measured by a thermocouple connected to a data acquisition instrument. A glass shield is used to avoid wind disturbance (Figure 6b). To reduce the thermal contact resistance, a silicon pad is applied between the heating plate and the sample. With the increasing heating power, both the equilibrium temperatures of the PCS and the foil increase, as shown in Figure 6c. With the same heating power, the temperature of the PCS is always lower than that of the Al foil owing to the thermal management by heat dissipation by radiation. For instance, with the heating power of 0.40 W/cm$^2$, the real temperature of the PCS sample (120.7 ℃) is 12.2 ℃ lower than that of the Al foil (132.9 ℃). As shown in Figure 6d, the cooling performance is contributed mainly by the emitted radiation within 5~8 $\mu$m and 8~14 $\mu$m. With the temperature increasing, the contribution of emitted power through 5~8 $\mu$m window will also increase. The demonstrated photonic crystal has a good cooling performance in high temperature.

# Conclusion

Consisting of ZnO/Ag/ZnO film perforated with a 2D periodic aperture array, the proposed photonic crystal achieves VICC with heat dissipation by radiation by utilizing EOT. With a high visible transmittance and low emittances in two atmospheric windows, the VICC performance of this PCS sample is demonstrated experimentally at high temperatures within 3~5 $\mu$m and 8~14 $\mu$m, respectively. With a heating power of 0.4 W/cm$^2$, the temperature of PCS is 12.2 ℃ lower than a low-emittance Al foil. The proposed VICC photonic crystal with the performance of thermal management provides a guideline on coordinated control of light and heat, indicating a potential application in energy & thermal technologies.

## Simulation, Fabrication and Characterization

### FDTD simulation

The radiative properties of the photonic crystal are analyzed numerically by FDTD method. By Eastwave software ( Eastwave Inc. ), the radiative properties in visible and infrared bands are calculated with the normally incident light polarized in x direction (Figure 1c). The geometric parameters are determined based on the required radiative properties in visible and infrared bands. To realize high visible transmittance and infrared reflectance, the thicknesses of ZnO/Ag/ZnO are 38.7/14.1/50.6 nm ($t_1$/$t_A$/$t_2$ in Figure 1c), respectively. To obtain EOT in 5~8 $\mu$m, the diameter of aperture $D$ is 3 $\mu$m and the period $P$ is 5.5 $\mu$m. The refractive index and extinction coefficient of Ag are expressed by the Lorentz-Drude model [62]. The optical constants of ZnO in visible and infrared are taken from Ref. [63] and Ref [64], respectively. In the directions of the distribution of the 2D array apertures (i.e., x and y), periodical boundary condition is applied. In the direction of the incident wave (i.e., z), the perfectly matched layer absorbing boundary condition is applied. In the simulation, a free-standing photonic crystal (FSPC) and the photonic crystal on a silica substrate (PCS) are considered with the refractive index of $SiO_2$ taken from Ref [65].

### Fabrication

In this work, the photonic crystal was prepared on a wafer silica substrate (diameter 2.5 cm, thickness 1.0 mm). The magnetron sputtering deposition technology (Kurt J. Lesker. LAB 18) was applied to deposit the films of ZnO and Ag in a high vacuum chamber with a pressure value of $4.8 \times 10^{-6}$ torr . The ZnO target used in the process is from Kurt J. Lesker (US) with a purity of 99.999% and a diameter of 3 inches, while the Ag target comes from Zhongnuo New Material technology Co. Ltd (China) with a purity of 99.99% and a diameter of 3 inches. For the deposition of ZnO, its growing speed is 0.5 Å/s with an operating power of 150 W and an air pressure of 3 mTorr. For the deposition of Ag, the growing speed is 5 Å/s with an operating power of 150 W and an air pressure of 5 mTorr. To fabricate cylindrical cavities in a square array, photolithography technology was applied. The photoresist was applied on the substrate by a spin coater and a hot plate (SUSS Lab spin6 and SUSS Delta HP8). The unfinished sample was covered with a Cr etch mask which was prepared by maskless lithograph system (ATD 1500) and exposed by an optical aligner (SUSS MA6). Then, the pattern was transferred on substrate by negative tone and magnetron sputtering was applied to deposit ZnO/Ag/ZnO film. Finally, by liftoff process with PMMA, the remaining photoresist was removed and the sample with a 2D aperture array was prepared.

**Characterization**

Through a scanning electron microscopy (SEM, SIRION200), the cross-section of the photonic crystal can be observed. For the measurement of normal incident radiative properties of the PCS sample in visible and infrared bands, a UV-visible-NIR spectrophotometers (SolidSpec-3700 DUV) and a Fourier transform infrared (FTIR) spectrometer (Bruker VERTEX 80) with an external integrating sphere were used, respectively.

**Acknowledges**

This work was funded by the National Natural Science Foundation (No. 51576188). This work was partially carried out at the USTC Center for Micro and Nanoscale Research and Fabrication, and we thank Yu Wei, Yizhao He, Wenjuan Li, Jinlan Peng, and Xiuxia Wang for their help with micro/nano fabrication. The authors give thanks to Hongxin Yao and Wei Yang for their help with the measurement of radiative properties. The authors also give special thanks to Zijun Wang and Fan Yang for their help with the idea and thermal experiment.

# Table captions

Table 1 The spectral emittance for several typical assumed structures.

Table 1 The spectral emittance for several typical assumed structures.

| Emittance | | Atmospheric windows | | Non-atmospheric window | Rest band |
|---|---|---|---|---|---|
| | | 3~5 $\mu$m | 8~14$\mu$m | 5~8 $\mu$m | |
| Structure I: Broadband zero-emittance | | 0 | 0 | 0 | 1 |
| Structure II: Ideal VICC for high temperature target | | 0 | 0 | 1 | 1 |
| Structure III | Low emittance | 0.05 | 0.05 | 0.05 | 1 |
| Structure IV | | 0.05 | 0.05 | 1 | 1 |

# Figure captions

Figure 1 Scheme of VICC photonic crystal for a high temperature target. **a)** Sketch of the scheme combining a photonic crystal, visible camouflage coating and a thermal insulator for high-temperature VICC. **b)** The emittance spectrum of structures with an ideal surface for high-temperature VICC (black solid line), emitted radiation by a 680 K blackbody (green solid line) and broadband zero-emittance surface (pink dash line) along with the atmospheric transmittance spectrum. **c)** Schematic of the photonic crystal: a ZnO/Ag/ZnO film perforated with 2D periodic array of apertures, with the diameter of aperture $D$, the period $P$ and the thicknesses of ZnO/Ag/ZnO being $t_1/t_A/t_2$.

Figure 2 **a)** Theoretical surface temperature $T_s$ reduced by structure I (broadband zero-emittance, dash line) and structure II (ideal VICC for high temperature target, solid line) with thermal insulator. Contributions to the infrared signal intensity reduction of thermal insulation (blue area), low emittance ($\varepsilon = 0.05$) (purple area) and high emittance in 5~8 $\mu$m (orange area) at different target temperatures $T_h$. **b)** in 3~5 $\mu$m, **c)** in 8~14 $\mu$m.

Figure 3 PCS sample and its radiative properties. **a)** Measured and simulated directional-hemispherical ($T_{dh}$) and directional-specular ($T_{ds}$) visible (0.38~0.78 $\mu$m) transmittance as well as the SEM of ZnO/Ag/ZnO. **b)** Sample photo. **c)** and **d)** The top-view and side-view SEM results of PCS sample. The radiative properties of FSPC and PCS. **e)** The simulated directional-hemispherical ($R_{dh}$) and directional-specular reflectance ($R_{ds}$) as well as $T_{dh}$ and $T_{ds}$ of FSPC within 3~15 $\mu$m. **f)** The simulated $R_{dh}$, $R_{ds}$ and the measured $R_{dh}$ of PCS in 3~15 $\mu$m

Figure 4 Electric fields at the cross-section of single unit for FSPC and PCS on x/z directions normalized by incident wave. For FSPC, **a)** $|E_x/E_0|$ and $|E_z/E_0|$ at $\lambda = 5.5$ $\mu$m, **b)** $|E_x/E_0|$ and $|E_z/E_0|$ at $\lambda = 6.1$ $\mu$m, **c)** $|E_x/E_0|$ and $|E_z/E_0|$ at $\lambda = 14.0$ $\mu$m. For PCS sample, **d)** $|E_x/E_0|$ and $|E_z/E_0|$ at $\lambda = 5.5$ $\mu$m, **e)** $|E_x/E_0|$ and $|E_z/E_0|$ at $\lambda = 6.4$ $\mu$m, **f)** $|E_x/E_0|$ and $|E_z/E_0|$ at $\lambda = 6.8$ $\mu$m, **g)** $|E_x/E_0|$ and $|E_z/E_0|$ at $\lambda = 14.0$ $\mu$m.

Figure 5 The demonstration of VICC of PCS within two atmospheric windows. **a)** The demonstration of visible camouflage. **b**) The platform for infrared camouflage demonstration. **c**) the infrared images within 3~5$\mu$m and 8~14$\mu$m infrared cameras at $T_s$= 74.2°C and $T_s$= 126.6 °C.

Figure 6 **a**) Measured and calculated radiation temperatures of the PCS sample as a function of the surface temperature, $T_s$. Dash lines represent the calculated radiation temperature for the FSPS. **b**) Schematic of cooling demonstration. **c**) The temperature rise with the growing heating power as a function of time. **d**) The comparison of the emitted powers by PCS sample within 3~5, 5~8 and 8~14 $\mu$m as a function of its temperature, $T$ (°C).

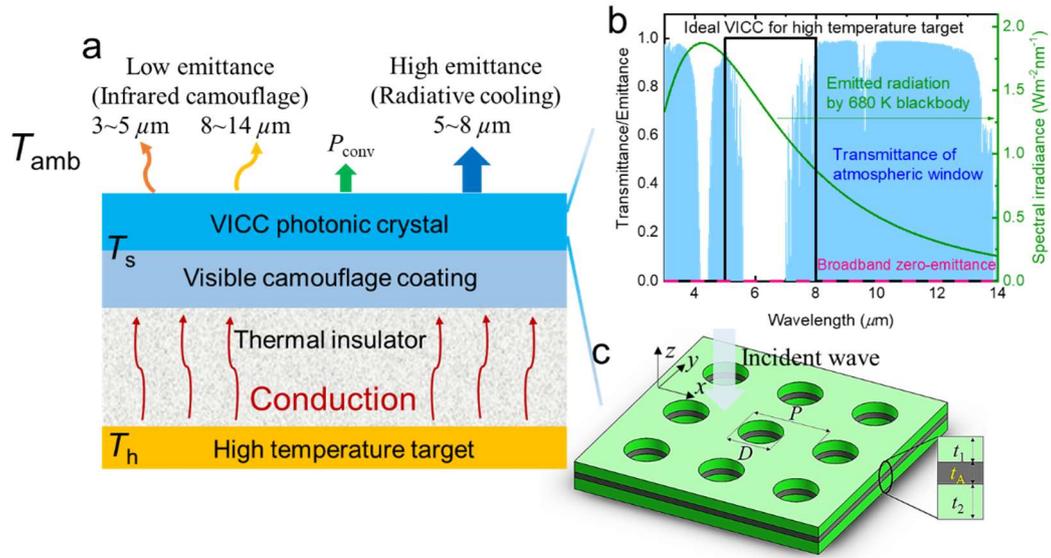

Figure 1 Scheme of VICC photonic crystal for a high temperature target. **a)** Sketch of the scheme combining a photonic crystal, visible camouflage coating and a thermal insulator for high-temperature VICC. **b)** The emittance spectrum of structures with an ideal surface for high-temperature VICC (black solid line), emitted radiation by a 680 K blackbody (green solid line) and broadband zero-emittance surface (pink dash line) along with the atmospheric transmittance spectrum. **c)** Schematic of the photonic crystal: a ZnO/Ag/ZnO film perforated with 2D periodic array of apertures, with the diameter of aperture $D$, the period $P$ and the thicknesses of ZnO/Ag/ZnO being $t_1/t_A/t_2$.

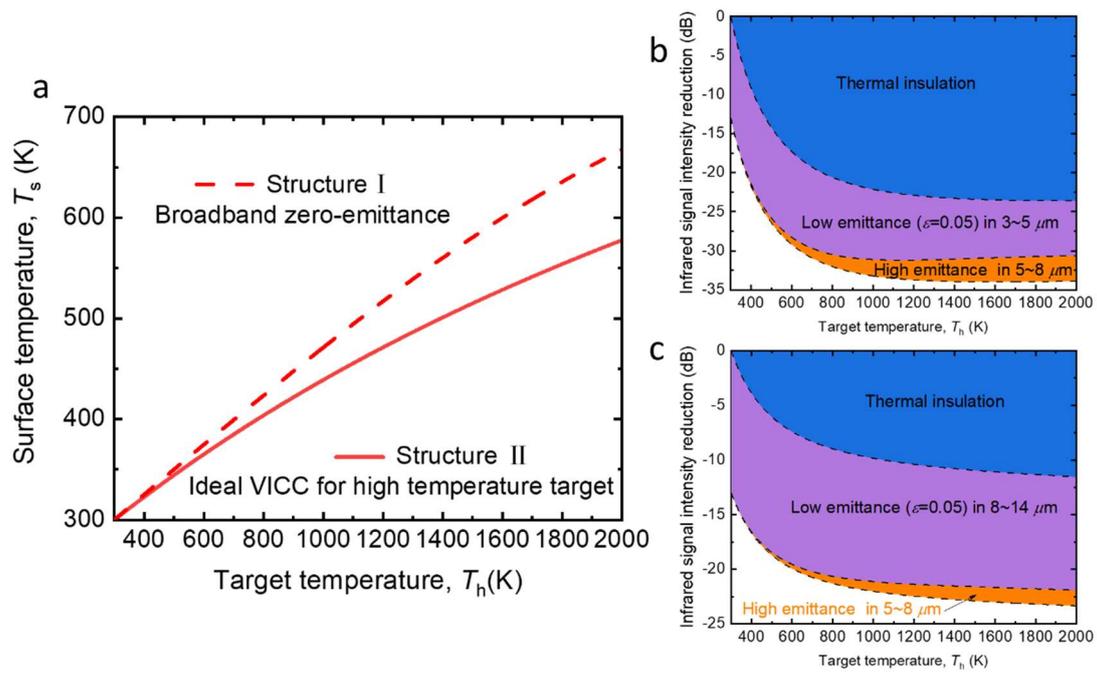

Figure 2 **a**) Theoretical surface temperature $T_s$ reduced by structure I (broadband zero-emittance, dash line) and structure II (ideal VICC for high temperature target, solid line) with thermal insulator. Contributions to the infrared signal intensity reduction of thermal insulation (blue area), low emittance ($\varepsilon$= 0.05) (purple area) and high emittance in 5~8$\mu$m (orange area) at different target temperatures $T_h$. **b**) in 3~5 $\mu$m, **c**) in 8~14 $\mu$m.

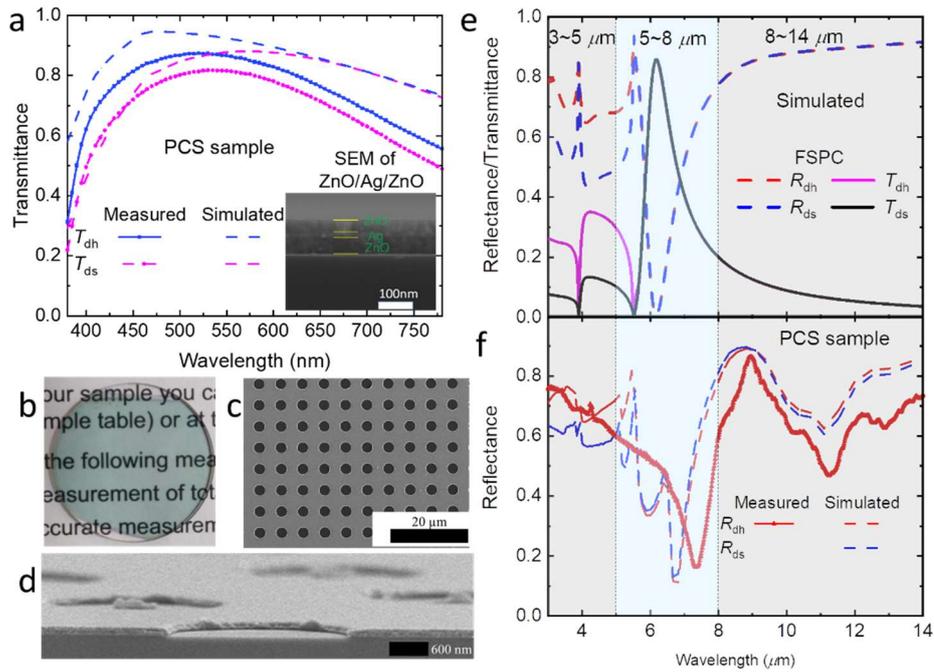

Figure 3 PCS sample and its radiative properties. **a**) Measured and simulated directional-hemispherical ($T_{dh}$) and directional-specular ($T_{ds}$) visible (0.38~0.78 $\mu$m) transmittance as well as the SEM of ZnO/Ag/ZnO. **b**) Sample photo. **c**) and **d**) The top-view and side-view SEM results of PCS sample. The radiative properties of FSPC and PCS. **e)** The simulated directional-hemispherical ($R_{dh}$) and directional-specular reflectance ($R_{ds}$) as well as $T_{dh}$ and $T_{ds}$ of FSPC within 3~15 $\mu$m. **f)** The simulated $R_{dh}$, $R_{ds}$ and the measured $R_{dh}$ of PCS in 3~15 $\mu$m.

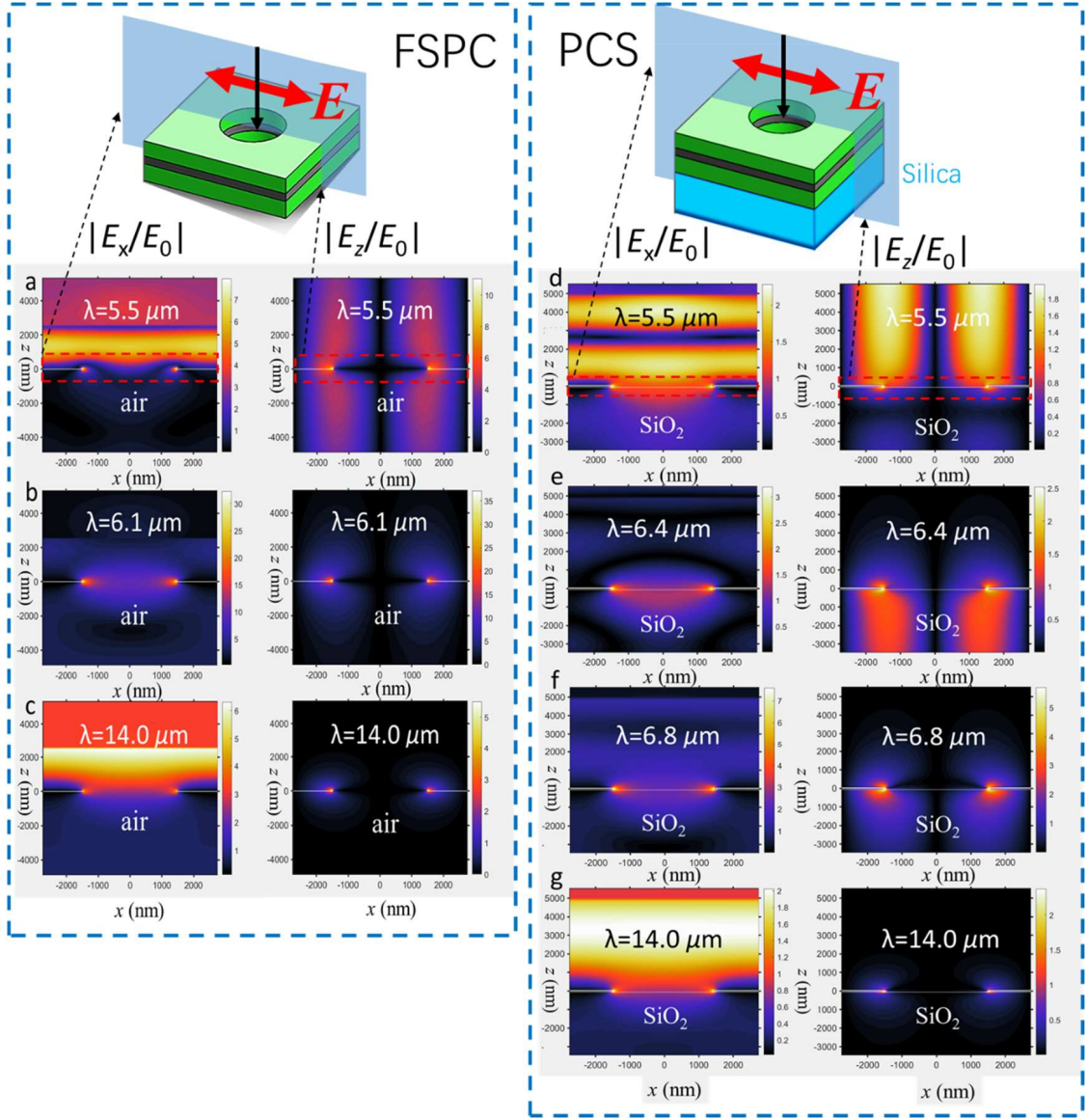

Figure 4 Electric fields at the cross-section of single unit for FSPC and PCS on x/z directions normalized by incident wave. For FSPC, **a)** $|E_x/E_0|$ and $|E_z/E_0|$ at $\lambda = 5.5\ \mu m$, **b)** $|E_x/E_0|$ and $|E_z/E_0|$ at $\lambda = 6.1\ \mu m$, **c)** $|E_x/E_0|$ and $|E_z/E_0|$ at $\lambda = 14.0\ \mu m$. For PCS sample, **d)** $|E_x/E_0|$ and $|E_z/E_0|$ at $\lambda = 5.5\ \mu m$, **e)** $|E_x/E_0|$ and $|E_z/E_0|$ at $\lambda = 6.4\ \mu m$, **f)** $|E_x/E_0|$ and $|E_z/E_0|$ at $\lambda = 6.8\ \mu m$, **g)** $|E_x/E_0|$ and $|E_z/E_0|$ at $\lambda = 14.0\ \mu m$.

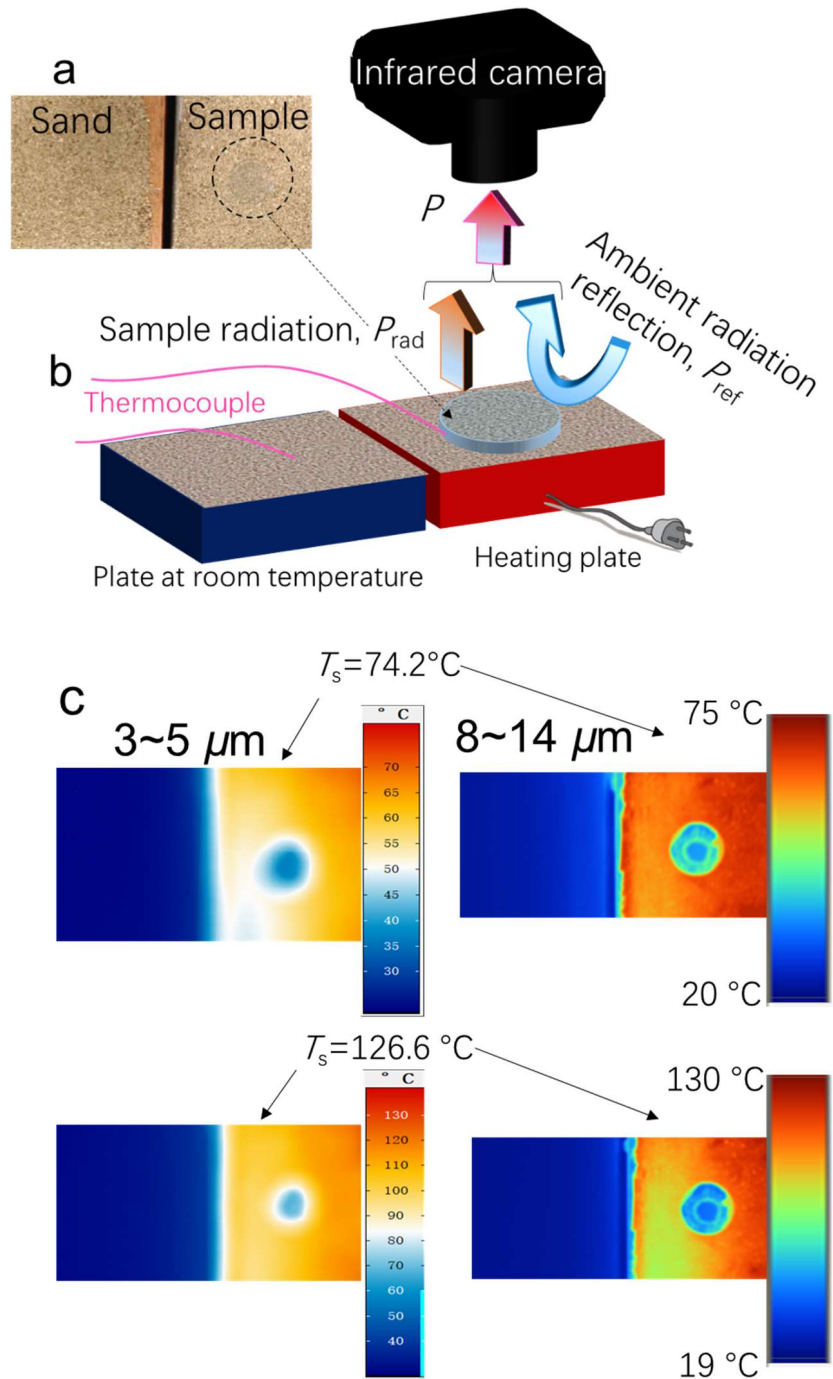

Figure 5 The demonstration of VICC of PCS within two atmospheric windows. **a**) The demonstration of visible camouflage. **b**) The platform for infrared camouflage demonstration. **c**) the infrared images within 3~5*μ*m and 8~14*μ*m infrared cameras at $T_s$= 74.2°C and $T_s$= 126.6 °C.

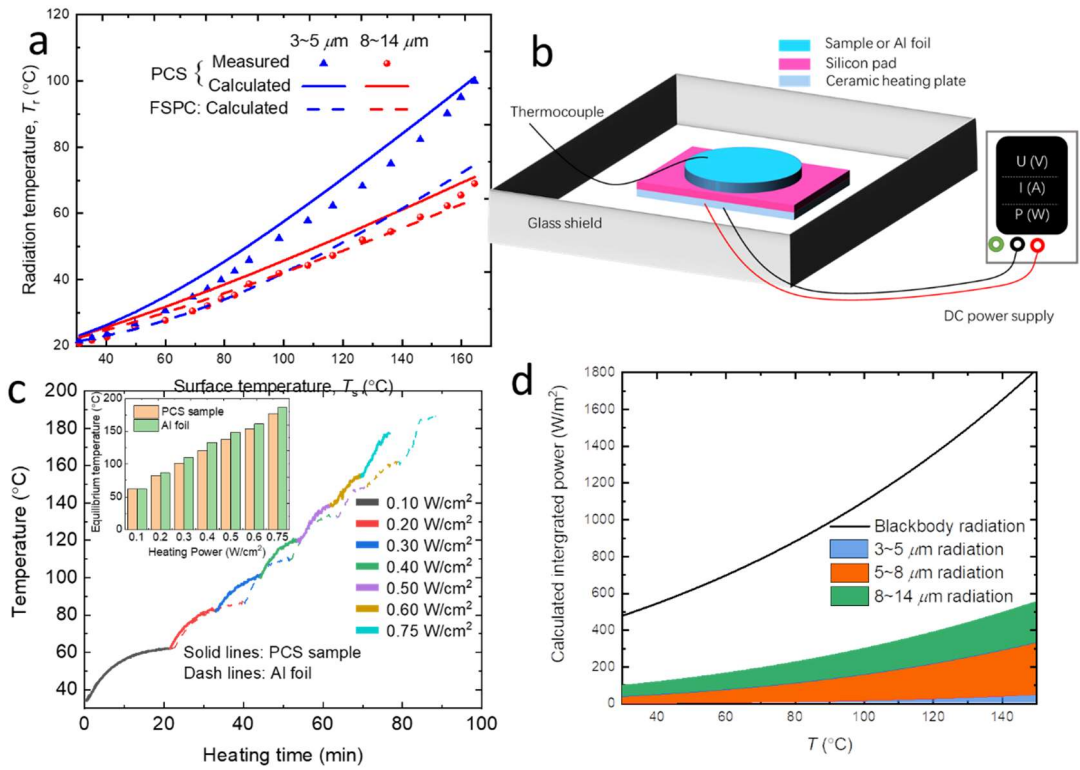

Figure 6 **a**) Measured and calculated radiation temperatures of the PCS sample as a function of the surface temperature, $T_s$. Dash lines represent the calculated radiation temperature for the FSPS. **b**) Schematic of cooling demonstration. **c**) The temperature rise with the growing heating power as a function of time. **d**) The comparison of the emitted powers by PCS sample within 3~5, 5~8 and 8~14 $\mu$m as a function of its temperature, $T$ (°C).